\documentclass[multphys,vecphys]{svmult}
\usepackage{graphicx}        
\usepackage{multicol}        
\usepackage[bottom]{footmisc}

\makeindex             

\begin{document}

\title*{A VIMOS-IFU survey of $z \sim 0.2$ massive lensing galaxy clusters:
constraining cosmography}
\titlerunning{A VIMOS-IFU survey of $z \sim 0.2$ massive lensing galaxy clusters} 
\author{Genevi\`eve Soucail\inst{1}\and
Giovanni Covone\inst{2,3}\and
Jean-Paul Kneib\inst{3}}
\institute{OMP, Laboratoire d'Astrophysique de Toulouse-Tarbes, 
14 Avenue Belin, 31400 Toulouse, France, 
\texttt{soucail@ast.obs-mip.fr}
\and INAF -- Osservatorio Astronomico di Capodimonte, Naples, Italy
\and OAMP, Laboratoire d'Astrophysique de Marseille, France
}
\maketitle

\begin{abstract}
We present an integral field spectroscopy survey of rich clusters
of galaxies aimed at studying their lensing properties.  Thanks to
knowledge of the spectroscopic caracteristics of more than three
families of multiple images in a single lens, one is able in principle
to derive constraints on the geometric cosmological parameters. We show
that this ambitious program is feasible and present some new results,
in particular the redshift measurement of the giant arc in A2667 and
the resdshift confirmation of the counter-image of the radial arc in
MS2137--23. Prospects for the future of such program are presented.
\end{abstract}

\section{Cosmological applications of gravitational lensing}

\begin{figure}
\centering
\includegraphics[height=4cm]{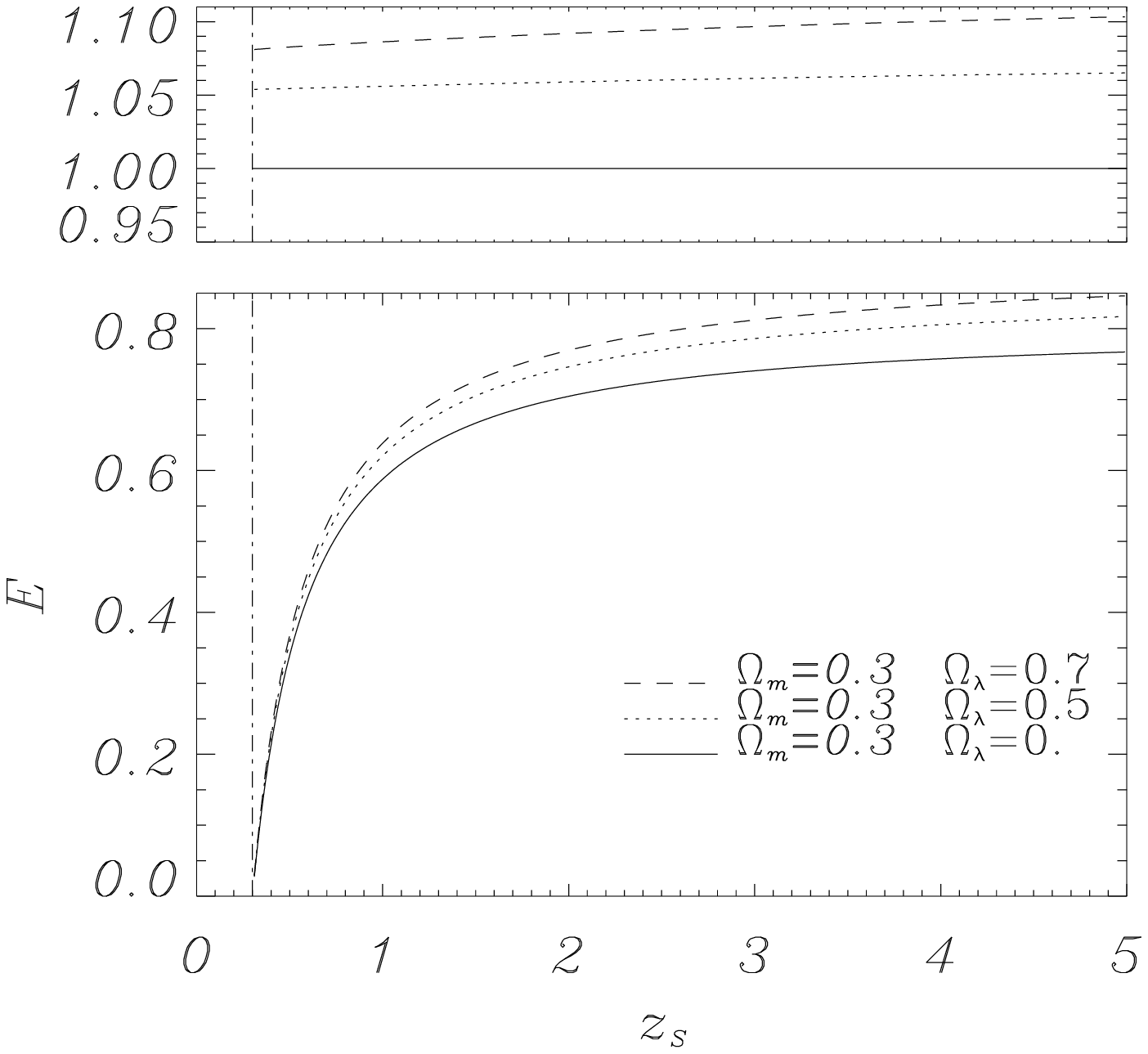}
\includegraphics[height=4cm]{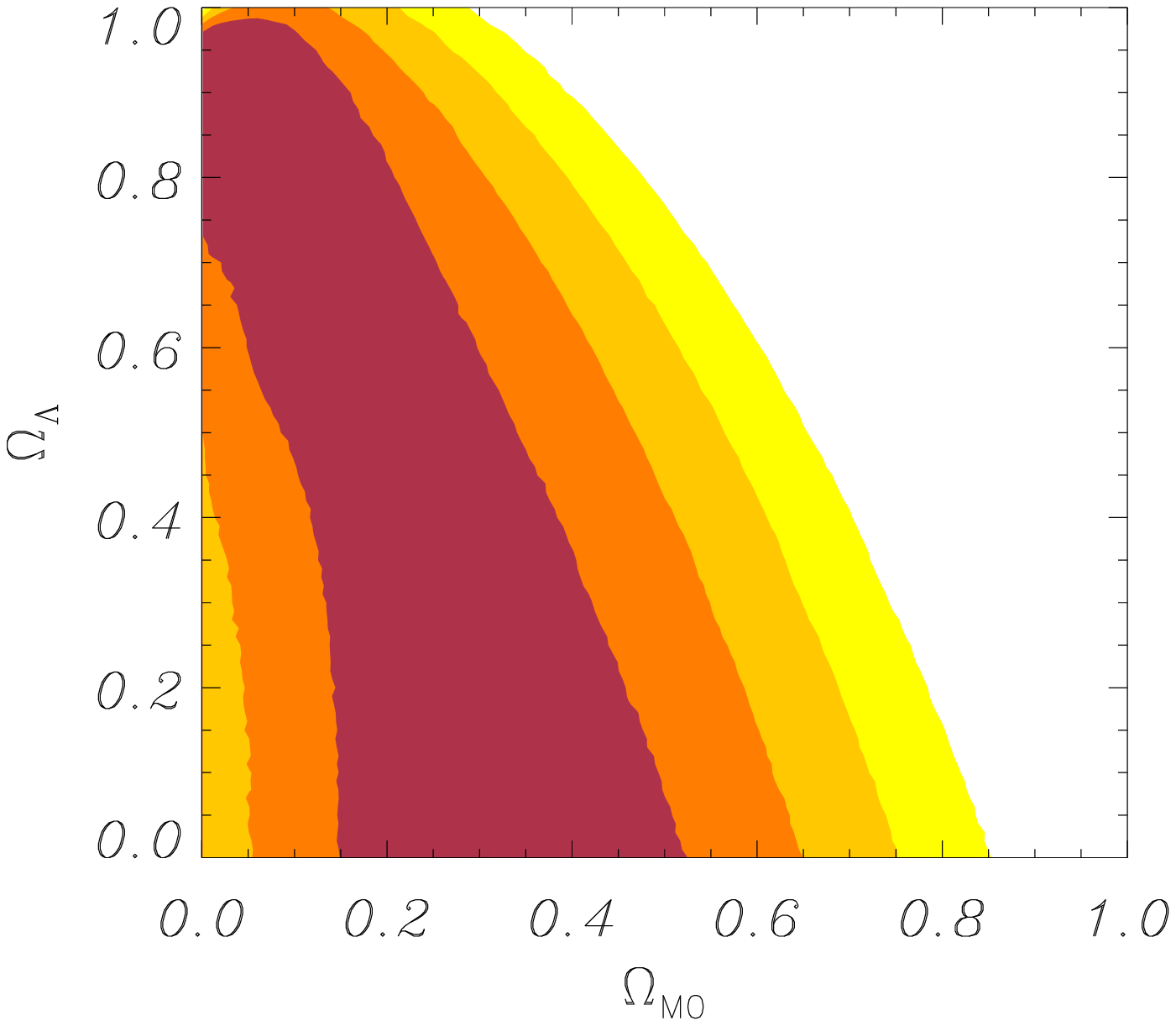}
\caption{Left: Evolution of the E-term with redshift for different
cosmological models, for a lens at $z=0.3$. Right: Confidence levels
obtained from the modeling of the cluster-lens Abell 2218
\cite{soucail04}}
\label{a2218}       
\end{figure}

Gravitational lensing has been deeply renewed since 15 years now, and many
extensive reviews are available in the litterature \cite{schneider92}. So
we simply remind here the basic lensing equation, which corresponds to
a mapping of the image plane on the source plane \[ \vec \beta = \vec
\theta - \vec \nabla \psi (\vec \theta) \qquad \qquad \psi (\vec \theta)
= \frac{2}{c^2} \frac{D_{LS}}{D_{OS}D_{OL}} \varphi (\vec \theta) \]
where $\varphi (\vec \theta)$ is the newtonian gravitational potential
of the deflecting lens. The distances are angular-distances between
the lens, the source and the observer. The Einstein radius, defined
as \[ R_E = \sqrt{\frac{4 G M}{c^2} \ \frac{D_{OL} D_{LS}}{D_{OS}}}
= 2.6'' \left(\frac{\sigma^2}{c^2}\right) \frac{D_{LS}}{D_{OS}} \]
represents the radius of the circular image in the ideal case of a
source perfectly aligned behind a point source of mass $M$ or behind
a singular isothermal sphere with velocity dispersion $\sigma$. For
any lensing mass, it represents the typical angular scale of the lensed
images. Numerically, it is about $1-3''$ for a galaxy lens and $10-30''$
for a cluster of galaxies. In practice, gravitational lensing has many
astrophysical applications, related to either the determination of the
lensing potential and the mass distribution in galaxies and in clusters of
galaxies, or the study of the distant galaxies thanks to the magnification
of their images \cite{mellier99}.  But it also depends on the geometrical
cosmological parameters and represents an alternative possibility
to constrain these parameters, independantly from other methods like
supernovae or CMB fluctuations. This method was explored recently in
details \cite{golse02}.  It requires several sets of multiple images
from different sources at different redshifts through the same lens.
For each source the lens equation can be written as \[ \vec \theta_S =
\vec \theta_{I} - \frac{2 \sigma_0^2}{c^2} \vec f(\vec \theta_I, \theta_C,
\alpha, \ldots ) \times E(\Omega_M, \Omega_\Lambda, z_L, z_S) \] where
the $f$ function includes all the caracteristics of the lensing potential
and $E=D_{LS}/D_{OS}$ includes the cosmological dependence [$(\Omega_M,
\Omega_\Lambda)$ or $(\Omega_M, w)$ for an euclidian universe] as well as
the lens and source redshifts. Provided $z_L$ and $z_S$ are well known for
two sets of multiple images, one can in principle solve the lens equation
for both sets and constrain the E-term, with a well-defined degeneracy
between the two geometrical cosmological parameters used \cite{golse02}.
However the cosmological dependence in the E-term is weak and better
constraints require more than two sets of multiple images, spread in
redshift as well as a very accurate mass determination in the lens
(Fig. \ref{a2218}). 

This method was first attemped in the cluster Abell 2218 ($z_L=0.17$,),
a strong gravitational lens close to the ideal case: it displays at least
4 sets of multiple images with redshifts ranging from $z_{S1}=0.702$
to $z_{S4}=5.576$. An accurate mass model was developped \cite{kneib96},
including 2 major mass components located on the two brightest galaxies,
as well as the contribution of about 17 galaxy masses scaled by their
luminosity.  This model prescription was constrained more accurately
by including the 4 families of multiple images, and the likelihood
distribution in the $(\Omega_M, \Omega_\Lambda)$ plane finally displayed
the expected degeneracy (Fig. \ref{a2218}), reinforcing the validity of
the mass model \cite{soucail04}.

\section{Observations of massive lensing clusters with VIMOS/IFU}
In order to tighten the likelihood distribution in the $(\Omega_M,
\Omega_\Lambda)$ plane we need to add more cluster-lenses with
similar numbers of multiple images but different lens redshifts or
mass distributions. This is the main motivation of a survey of massive
cluster lenses started two years ago at ESO and using the IFU mode of
VIMOS on the VLT. This instrument configuration has many advantages,
thanks to its wide field of view of $1'\times 1'$ in order to get as
many redshifts as possible, both on cluster galaxies and on lensed arcs
with curved shapes. The high density of objects in the central cores of
clusters is therefore mapped in a single shot, allowing both to study
the dynamics of the cluster galaxies and the redshift determination of
the brightest multiple arcs.

\begin{table}
\centering
\caption{Summary of the VIMOS/IFU observations of the survey}
\label{tab_survey}       
\begin{tabular}{lcccc}
\hline\noalign{\smallskip}
Cluster & redshift \, & \# pointings & Run & $T_{exp}$ (ksec)  \\
\noalign{\smallskip}\hline\noalign{\smallskip}
Abell 1689 & 0.184 & 4 & May-June 2003 & 59.4 \\
Abell 2390 & 0.233 & 3 & June 2003 & 32.4 \\
AC 114 & 0.312 & 2 & June 2004 & 21.6 \\
Abell 2667 & 0.233 & 1 & June 2003 & 10.8  \\
MS2137--23 & 0.310 & 1 & June 2003 & 10.8 \\
Abell 68 & 0.255 & 1 & August 2004 & 5.4 \\
\noalign{\smallskip}\hline
\end{tabular}
\end{table}

Up to now, 6 clusters have been observed in low resolution mode ($R
\sim 200$), with a usefull spectral range from 4000 to 6800 \AA 
(Table \ref{tab_survey}). Most clusters have a
redshift in the range 0.2--0.3 which is optimal for our 
programue. Data reduction was rather tricky, and is presented in
details in \cite{covone06}. 

\subsection{Abell 2667}
\begin{figure}
\centering
\includegraphics[height=6cm]{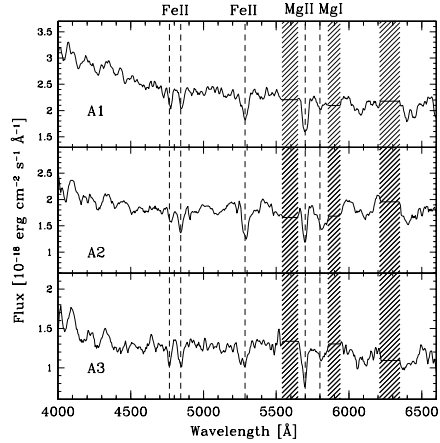}
\caption{Spectra of the three sub-images of the giant arc in A2667,
with line identifications at redshift $z=1.03$. }
\label{a2667_sp}       
\end{figure}

Abell 2667 was observed with IFU/VIMOS in a single pointing, splitted in
4 exposures of 45 mn each \cite{covone06} . It is a strong X-ray emitter
and multi-color HST imaging revealed a very prominent and bright luminous
arc corresponding to 3 merging images. Its spectrum was extracted and
splitted in 3 for each sub-image. The redshift confirmation is obvious
in this case (Fig. \ref{a2667_sp}). A few very faint multiple images
candidates were also identifed. However, by exploring the final datacube,
we were not able to determine more new arc redshifts, preventing the use
of A2667 for cosmological purposes. But with the present data, we were
able to determine the redshift of more than 20 cluster members, so a first
value of the velocity dispersion could be used for the lens modeling. A
detailed lens model was therefore built and the mass distribution compared
with the mass deduced from X-ray data and the virial mass \cite{covone06}
(see also Covone's talk, these proceedings).

\subsection{MS2137--23}
\begin{figure}
\centering
\includegraphics[height=4cm]{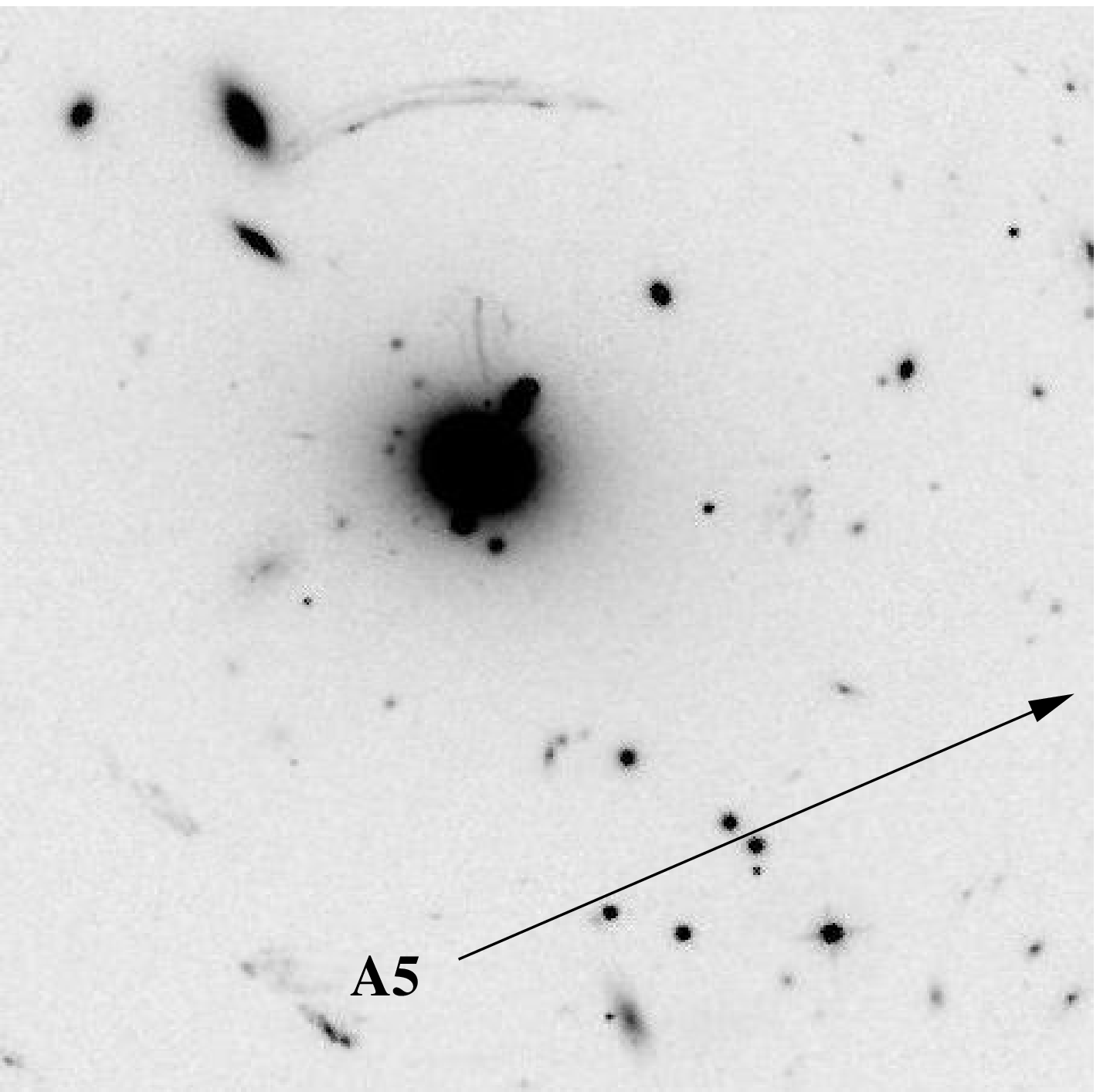}
\includegraphics[height=4cm]{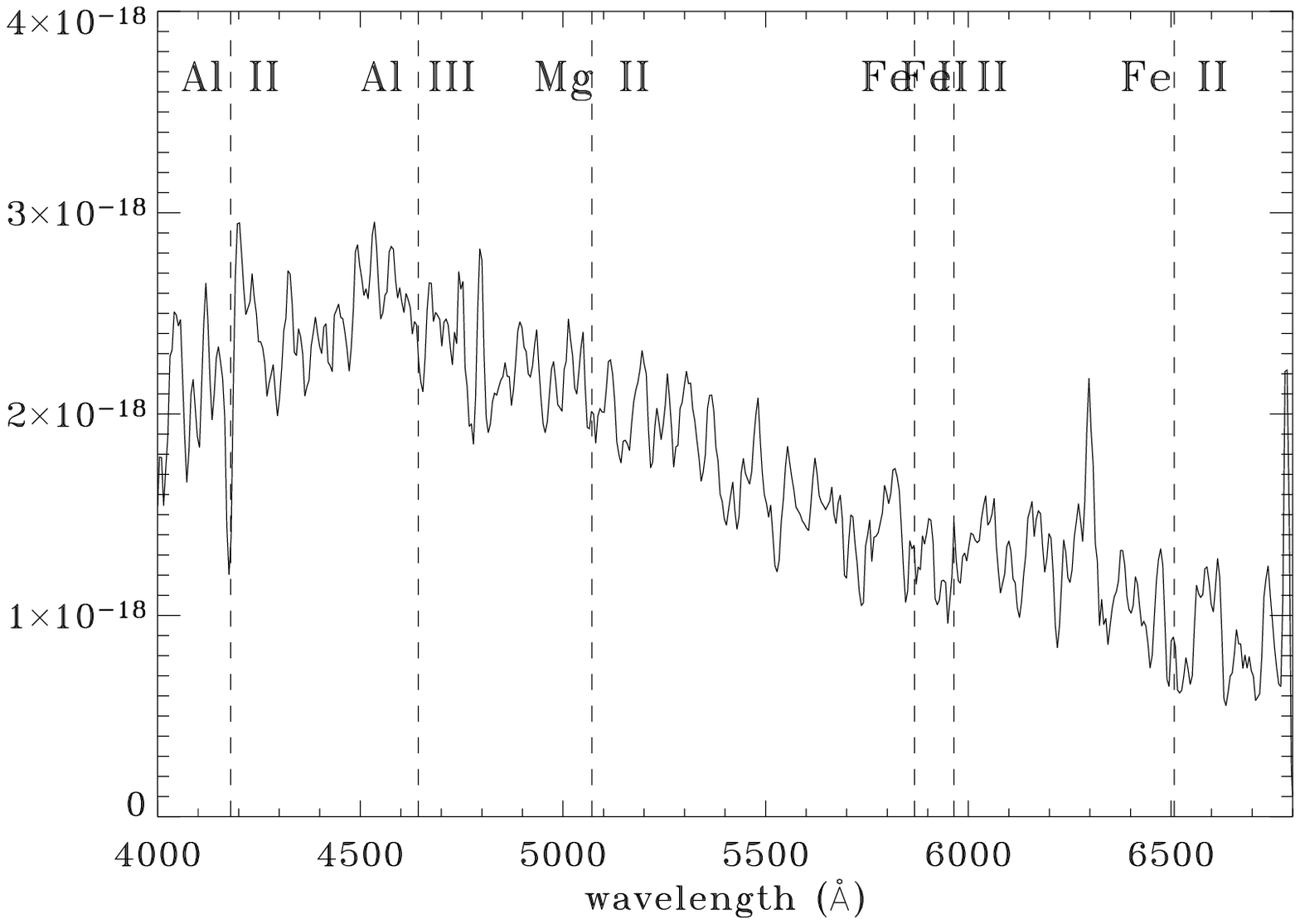}
\caption{Integrated spectrum of the counter-image of the radial arc in
MS2137--23, and the line identification with redshift $z=1.503$.}
\label{ms2137}       
\end{figure}

MS2137--23 is a well known gravitational lens with two independant systems
of lensed images: one tangential arc with its two counter-images and one
radial arc with its counter-image candidate. The two main arcs have been
observed spectroscopically with a redshift measurement at a redshift
$z \simeq 1.5$ for both systems \cite{sand02}.  Unfortunately this is
not favorable for cosmography although this lens is studied in details
in order to constrain the properties of the mass distribution in the
very center of clusters \cite{gavazzi05}.  However, from the datacube
built from our observations (2 exposures of 45mn each), we were able
to extract the spectrum of the counter-image of the radial arc (A5),
confirming its redshift at $z=1.503$ (Fig. \ref{ms2137}). This is the first
redshift confirmation that A5 is the counter-image of the radial arc.

\subsection{Abell 1689: the ultimate lens !}
Fortunately, there exits presently at least one cluster-lens which
deserves great attention because of its unprecedented number of arcs
and arclets. From deep HST/ACS observations of Abell 1689, more than 30
systems of multiple images are identified in the center 
\cite{broadhurst05}. The lens
has a complex mass distribution and its lens modeling is not easy. In
addition only a few multiple images presently have measured redshifts,
because of their faintness. We have started an VIMOS/IFU survey of this
lens in 2003, when the instrument was not completely
stabilized. This prevented an accurate data reduction, although we still
hope to extract some exciting results from the datacubes. However,
several spectroscopic redshifts are already available, in particular
from previous long slit observations and we should be able to
provide a good lens model as well as some new constraints on
cosmography. Note that a preliminary attempt \cite{broadhurst05} did not 
give interesting constraints, partly because they used 
photometric redshifts only, for most of the arcs. The key point will be
clearly to include additive contraints on the lens model either from 
weak lensing measurements at large distance or from X-ray data.

\section{Conclusions and prospects}
We have started an ambitious observing programme, well 
suited to 3D spectroscopy and aimed at studying in details a sample of
strong gravitational lenses. The main difficulty is the
faintness of the arcs which requires to push the present
spectrographs to their limits. The results are therefore very sensitive 
to the data quality at the output of the telescope and to the
data reduction which must be done with great care. These
fundamental steps are in progress and we have demonstrated that faint
objects 3D spectroscopy is feasible in rich environments like clusters
of galaxies. Several scientific outputs from these
data are presented, others are in Covone's
talk (these proceedings) concerning the properties of the cluster
galaxies. 
%
%


\end{document}